\documentclass[12pt]{article}\usepackage[]{graphicx}\usepackage[]{xcolor}
\makeatletter
\def\maxwidth{ %
  \ifdim\Gin@nat@width>\linewidth
    \linewidth
  \else
    \Gin@nat@width
  \fi
}
\makeatother

\definecolor{fgcolor}{rgb}{0.251, 0.251, 0.282}

\usepackage{framed}
\makeatletter
 {\par\unskip\endMakeFramed%
 \at@end@of@kframe}
\makeatother

\definecolor{shadecolor}{rgb}{.97, .97, .97}
\definecolor{messagecolor}{rgb}{0, 0, 0}
\definecolor{warningcolor}{rgb}{1, 0, 1}
\definecolor{errorcolor}{rgb}{1, 0, 0}
\newenvironment{knitrout}{}{} 

\usepackage{alltt}
\usepackage[colorlinks = true,linkcolor = blue, citecolor = blue]{hyperref}
\usepackage{bbm}

\def\ab{\textbf{a}}
\def\R{\mathbb{R}}
\def\N{\mathbb{N}}
\def\betag{\boldsymbol{\beta}} 
\def\deltag{\boldsymbol{\delta}} 
\def\pig{\boldsymbol{\pi}} 
\def\Ab{{\bf A}}

\def\Gb{{\bf G}}
\def\Cb{{\bf C}}
\def\Db{{\bf D}}
\def\Wb{{\bf W}}
\def\ab{{\bf a}}

\def\xb{{\bf x}}

\def\zb{{\bf z}}
\def\Ab{{\bf A}}

\def\Cb{{\bf C}}
\def\1b{{\bf 1}}

\def\E{{\rm E}}

\usepackage[margin=1.3in]{geometry}

\usepackage{float}
\newfloat{algorithm}

\usepackage{setspace}
\usepackage{bm,natbib,amsmath,amsthm,amsfonts}

\usepackage{algorithm}
\usepackage{algpseudocode}
\usepackage{tikz}
\usepackage{booktabs}
\usepackage{xcolor}
\usepackage{soul}

\title{Sequential Spatially Balanced Sampling}

\author{ Rapha\"el Jauslin,~ Bardia Panahbehagh~ and Yves Till\'e\footnote{Rapha\"el Jauslin and Yves Till\'e, Institute of statistics, University of Neuchatel, Av. de Bellevaux 51, 2000 Neuchatel, Switzerland, Email: raphael.jauslin@unine.ch; Bardia Panahbehagh, Department of Mathematics, Kharazmi University, Tehran, Iran.}
}

\IfFileExists{upquote.sty}{\usepackage{upquote}}{}

\def\ab{\textbf{a}}
\def\xb{\textbf{x}}
\def\zb{\textbf{z}}

\def\pib{\boldsymbol{\pi}} 
\IfFileExists{upquote.sty}{\usepackage{upquote}}{}
\begin{document}

\maketitle

\begin{abstract}
	Sequential sampling occurs when the entire population is unknown in advance and data are received one by one or in groups of units. This manuscript proposes a new algorithm to sequentially select a balanced sample. The algorithm respects equal and unequal inclusion probabilities. The method can also be used to select a spatially balanced sample if the population of interest contains spatial coordinates. \textcolor{black}{A simulation study is proposed and the results show that the proposed method outperforms other methods.}
	\\ \\\textbf{Keywords} inclusion probability, spread sampling, stream sampling, survey methods
\end{abstract}

\clearpage
\section{Introduction}\label{sec:intro}

A sample is said to be balanced if the Horvitz-Thompson estimators of the totals calculated from the sample are equal or nearly equal to the population totals. The use of balancing constraints was first proposed by \citet{gin:gal:29}. They selected a sample of 29 districts out of 214 to replicate certain population means. The selection of the sample was not random and the method was strongly criticized by Jerzy Neyman. It is now well known that a sample can be selected randomly and balanced simultaneously.  The cube method, which randomly selects a balanced sample, has been proposed by \citet{dev:til:04a}. Recently, \citet{leu:eus:jau:til:2021} proposed an improvement. They showed that changing the order of the units before running the algorithm can significantly increase the quality of the sample balance.

In environmental studies particularly, data contains spatial coordinates. When the data are spatially autocorrelated, it is often more accurate to spread the sample in space. Moreover, well-spread samples in space are  balanced on auxiliary variables \citep[see][]{graf:lund:13} even if the target parameters are nonlinear in the auxiliary variables. Many sampling methods are currently used to select a well-spread sample. A well-known algorithm is the generalized random tessellation sampling proposed by \citet{Stev:Olse:spat:2004}. It maps a multi-dimensional space into a real line and uses one dimension systematic sampling to select a sample. \cite{gra:lun:sch:12} proposed the local pivotal method that introduces repulsion between two nearby units to ensure that close units are not both selected. \textcolor{black}{\cite{ben:pier:17} proposed to recursively modify the inclusion probability vector using the within sample distance of the spatial structure. \citet{rob:mcd:pri:brown:18} proposed to use the properties of the Halton sequence to select a well-spread sample.} More recently, \citet{jauslin2020spatial} proposed adaptation of the cube method using a contiguity matrix to select a well-spread sample.

\textcolor{black}{\citet{graf:lund:13} showed that well-spread samples could also be balanced on auxiliary variables.}  But some methods can even do better. A combination of the cube method and the local pivotal method, named the doubly balanced sampling, has been proposed by \citet{gra:til:13}. It allows the selection of a sample, which is simultaneously well spread and balanced on auxiliary variables. They showed, moreover, that this method diminishes drastically the variance of the Horvitz-Thompson estimator. \citet{vall:till:2015} presented besides an application of the doubly balanced sampling design for forest inventories. Throu\-ghout this manuscript, we write ``spatially balanced sample'' or ``well-spread sample'' to refer to the same interpretation.

All these methods apply to finite populations. This means that we must have access to the entire population before using the sampling algorithm. We speak of streaming data when the data arrive one by one or in groups of units. \textcolor{black}{Examples are Internet network data, financial data, and environmental or biological studies over a long period of time. In these cases, the units may be distributed asymmetrically, meaning that unusual units may appear in the stream and have a significant impact on the estimator. Choosing balanced samples improves the precision of the estimator. Indeed, if an eccentric unit is selected, the balancing constraints ensure that the variance of the estimator is controlled. Having a sequential procedure for selecting a balanced sample could be useful if the data stream is huge.
Other methods that select a balanced sample need the entire population to select a sample, whereas with the proposed method only a group of units needs to be known. The reader may find it disturbing that the data set we choose in this manuscript is not in a stream. Indeed, we have chosen to test the method on a finite population for several reasons. First, we believe it is more important to compare methods on comparable data sets. We can then measure the performance of the method against the usual methods. In addition, this method is the only one known that selects a balanced sample sequentially, so we would have no point of comparison. Nevertheless, for us, this does not have an impact on the results. The simulation results show that the finite population method with sequential implementation has better properties in terms of variance reduction than the usual methods.} 

In this manuscript, we propose a new method to select a balanced sample and a spatially balanced if spatial coordinates are available. Moreover, the method is sequential, i.e., the algorithm does not need to get access to the entire population to run. In section~\ref{sec:notation}, we introduce the basic concept of survey sampling theory. In section~\ref{sec:balanced}, we expose the balancing equations formally and give insight into the interest of selecting a sample at the same time balanced and well spread. In section~\ref{sec:spatial}, we introduce spread measure while in section~\ref{sec:method} we present the outline of the method and explain how the method selects a spatially balanced sample. section~\ref{sec:simu} contains simulation results on two datasets.

{\color{black}
\section{Notation}\label{sec:notation}
Throughout this manuscript we consider a finite population $U$. The size of the population $N$ can be known in advance or estimated, especially in stream data, it is often the case that the population size is unknown. A sample $s\subset U$ is a subset of the population $U$. A sampling design is defined by a probability function $p(.)$ on all possible samples such that
$$
p(s) \geq 0 \text{ and }\sum_{ s\subset U}p(s) = 1.
$$
Let $S\subset U$ denote a random sample, a random variable with probability distribution defined by the sampling design $P(S = s) = p(s)$. Theoretically, inclusion probability $\pi_k \in (0,1)$, $k \in U$ can be deduced from the sampling design
\begin{equation*}\label{eq:pik}
	\pi_k = \textrm{P}(k \in S) = \sum_{s\subset U | k \in s} p(s).
\end{equation*}
In pratice, inclusion probabilities are predetermined by statisticians, theses could be equal with a fixed sample size, i.e. $\pi_k = n/N$, or unequal, for example proportional to an auxiliary variable.  Let $\pib=(\pi_1,\dots,\pi_N)^\top$ denote the vector of inclusion probabilities and $\deltag = (\delta_1,\delta_2,\dots,\delta_N)^\top$ the sample where
$$
\delta_k = \left\{\begin{array}{lll} 1 & \text{ if } k\in S\\ 0 & \text{ otherwise} . \end{array} \right.
$$
Let $y$ denote a variable of interest where $y_k$ denotes the value of the variable for a particular unit $k\in U$. Let $Y$ denote the total of the variable on the population $U$, i.e.,
$$
Y=\sum_{k\in U} y_k.
$$
It can be estimated unbiasedly by using the Horvitz-Thompson estimator of the total defined by
\begin{equation}\label{eq:HT}
	\widehat{Y}_{HT} = \sum_{k\in S} \frac{y_k}{\pi_k}= \sum_{k\in U} \frac{y_k \delta_k}{\pi_k}.
\end{equation}

If data are coming in a stream, we suppose that the units $k \in U$ are coming one by one or by groups of units, for example, subpopulations $\{U_1,U_2,U_3,\dots\}$. A necessary condition to sample in a stream data is to have a sequential sampling method, i.e., the algorithm takes the decision to select the unit in the sample or not and passes it to the next one. Indeed, as the size of the population might be unknown it is generally impossible to wait for the full population. Stream data are various, the reader may find a general framework in \citet{til:19a}.

Suppose $\xb_k = (x_{k1},x_{k2},\dots,x_{kp})^\top \in \R^p$ be auxiliary variables available for the unit $k\in U$. In an unequal probability sampling design, the aim is to select a sample with a fixed sample size proportional to the variable of interest $y$. Since the variable of interest is generally unknown for the whole population, we use an auxiliary variable, which has a strong correlation with the variable of interest, to establish the unequal inclusion probabilities. In the data stream context, the question of sample size with prescribed inclusion probabilities is notably discussed by \citet{cohen2009stream}. Inclusion probabilities can then be set up directly during the stream in order to reach the fixed sample size.

In addition to the auxiliary variables, let's suppose that the population contains spatial coordinates $\zb_k\in \R^q$. Population with spatial coordinates are often spatially correlated \citep{Wang20121}. In that case, selecting a well-spread sample will reduce the variance of the Horvitz-Thompson estimator \citep{graf:lund:13}. A stronger insight into this is presented in the next section.
}

\section{Balanced Sampling}\label{sec:balanced}

A sample $S$ is said to be balanced on auxiliary variables $\xb_k$, $k\in U$, if it satisfies the balancing equations given by
$$
\sum_{k\in S} \frac{\xb_k}{\pi_k} = \sum_{k\in U} \xb_k.
$$
Let $\Ab = (\xb_1/\pi_1,\xb_2/\pi_2,\dots,\xb_N/\pi_N)^\top = (\ab_1,\ab_2,\dots,\ab_N)^\top$ denote the auxiliary variables expanded by the inclusion probabilities. The selection of a balanced sample can be written as the linear program
$$
\left\{\begin{array}{lll}
	\Ab^\top\deltag = \Ab^\top\pib,\\
	\deltag\in \{0,1\}^N.
\end{array}\right.
$$

{\color{black}
\citet{dev:til:04a} have proposed the cube method to select a balanced sample. The method first consists in performing a random walk inside the hypercube to approach a balanced sample. This first step is called the flight phase. \citet{cha:til:06} have proposed a fast implementation, which modifies the inclusion probabilities $\pig$ into a slightly different vector of inclusion probabilities $\widetilde{\pig}$. This updated vector of inclusion probabilities $\widetilde{\pig}$ verifies exactly the balancing constraints $\Ab^\top\widetilde{\pig} = \Ab^\top\pig $. Moreover, it remains only at most $p$ units that have inclusion probabilities not equal to 0 or 1. The vector $\widetilde{\pig}$ is almost a sample as it remains only at most $p$ entries that are not equal to 0 or 1. Mathematically we have that $0 \leq \widetilde{\pi}_k \leq 1,~\E(\widetilde{\pi}_k) = \pi_k,~ k \in U$, and 
$$
\sum_{k\in S} \frac{\widetilde{\pi}_k\xb_k}{\pi_k} = \sum_{k\in U} \xb_k.
$$

A second phase is then launched on this updated vector to obtain a sample. This phase is called the landing phase and consist, either of dropping balancing constraint one by one or using a linear program to find out the best solution. The latter method can only be launched if the value of $p$ is not too high as it could lead to a combinatorial explosion. In the end, the random sample almost satisfies the constraint in the sense that
$$
\sum_{k\in S} \frac{\xb_k}{\pi_k} \approx \sum_{k\in U} \xb_k.
$$

}

\textcolor{black}{In presence of not only auxiliary variables but also spatial coordinates, \citet{gra:til:13} proposed a doubly balanced method that selects  samples which are well spread and balanced on auxiliary variables at the same time. Indeed, if the population of interest has spatial coordinates, the units are generally spatially correlated. In fact, if we assume a simple model, we can see that applying both approaches reduces the variance of the estimator.} Let suppose a general linear model with spatial correlation:
$$
y_k = \xb_k^\top\boldsymbol{\beta} + \varepsilon_k, \text{ for all } k \in U,
$$
where $\varepsilon_k$ is a random variable that satisfies $\E_M( \varepsilon_k) = 0$ and $\text{var}_M(\varepsilon_k) = \sigma^2(\xb_k) = \sigma_k^2$, with $\sigma^2(.)$ a Lipschitz continuous function and $\E_M(.),\text{var}_M$ are respectively the expectation and the variance under the model. Spatial correlation is modelled by the function
$$ \text{cov}_M(\varepsilon_k,\varepsilon_\ell) = \sigma_k\sigma_\ell \rho_{k\ell}, \text{ with } k\neq \ell \in U $$
where $ \rho_{k\ell}$ is a function that decreases when the distance between $k$ and $\ell$ increases. Under this model, \citet{gra:til:13} show that the anticipated variance of the Horvitz-Thompson estimator is
\begin{equation}\label{eq:spatjust} \E_p\E_M( \widehat{Y}_{HT} - Y)^2 = \E_p\left\{\left(\sum_{k\in S}\frac{\xb_k}{\pi_k} -\sum_{k\in U}\xb_k \right)^\top\boldsymbol{\beta} \right\}^2 + \sum_{k\in U}\sum_{\ell\in U} \sigma_k\sigma_\ell \rho_{k\ell} \frac{\pi_{k\ell} - \pi_k\pi_\ell}{\pi_k\pi_\ell},
\end{equation}
where $\pi_{k\ell} = \E_p(\delta_k\delta_\ell)$ are the joint inclusion probabilities and $\E_p(.)$ is the expectation under the design. From the first term of Equation~\eqref{eq:spatjust}, the reduction of the variance is done by selecting a balanced sample. Whereas the second term is reduced if the inclusion probabilities $\pi_{k\ell}$ is small while $\rho_{k\ell}$ is large. This means that a sample must be selected in a spread and balanced manner to minimize the anticipated variance.
	

{\color{black}
\section{Spreading Measures}\label{sec:spatial}
 A question arises naturally in spatial sampling: is there measures to see whether a sample is well spread?
\citet{Stev:Olse:vari:2003} proposed an index based on the Vorono\"i polygon. Let $b_i$, $i\in S$, be the sum of inclusion probabilities within the $i$th Vorono\"i polygons. They showed that the expected value of $b_i$ is equal to 1 and proposed to measure how vary these sums using the following quantity:
\begin{equation}\label{eq:sb}
   B = \frac{1}{n}\sum_{i\in S} (b_i - 1)^2.
\end{equation}

\citet{moran1950notes} proposes to measure spatial correlations in particular case where the spatial coordinates are placed on a grid. \citet{til:dic:esp:giu:18, jauslin2020spatial} propose a normalized version that uses the notion of a spatial weight matrix $\Wb$. Readers can find more information on how the spatial weight matrix is computed in \cite{jauslin2020spatial}. Moran's index $I$ is given by
\begin{equation}\label{eq:moran}
	I = \frac{(\deltag - \bar{\deltag}_w)^\top \Wb (\deltag - \bar{\deltag}_w) }{\sqrt{(\deltag - \bar{\deltag}_w)^\top \Db (\deltag - \bar{\deltag}_w)(\deltag - \bar{\deltag}_w)^\top \Gb (\deltag - \bar{\deltag}_w) }},
\end{equation}
where $\deltag$  is the sample and
$$
\bar{\deltag}_w = \frac{\deltag^\top \Wb \1b}{\1b^\top \Wb \1b},
$$
$\Db$ is the $N\times N$ diagonal matrix containing for each $k$, $\sum_{\ell\in U} w_{k\ell} $ on its diagonal,
$$
\Gb = \Cb^\top\Db\Cb,~~~~ \Cb = \Db^{-1}\Wb - \frac{\1b\1b^\top\Wb}{\1b^\top\Wb\1b},
$$
and $\1b$ is a column vector of $N$ ones. These two measures are the ones used to measure the spread in Section \ref{sec:simu}.

}


\section{Outline of the Proposed Method}\label{sec:method}
In this section, we present the main idea of the method. Suppose a population $U$ where units arrive sequentially, by groups of units or one by one. \textcolor{black}{Inclusion probabilities are supposed to be predetermined by the statistician. They are usually defined as unequal and proportional to an auxiliary variable. In this case, the fixed sample size is satisfied but it may be unknown depending on whether the population size is known. Inclusion probabilities can also be set up equal. In both cases, the following method will respect equal and unequal inclusion probabilities.} Additionally to the inclusion probabilities, we want to respect balancing equations and make a decision for the current unit observed in the stream. For simplicity, let's suppose that the current unit is the first one. \textcolor{black}{The algorithm works in the following way: it will wait for a certain number of units, denoted $J$, so that it can decide on the considered unit while satisfying the balancing equations. The number of units to wait depends on the number of auxiliary variables and the inclusion probabilities. The variable $J$ is tested at each step to see if a solution exists. The variable $J$ is therefore dynamic and changes after each decision. In general, $J$ is much smaller than $N$ and the algorithm compensates the decision taken on the current unit to the $J-1$ remaining units.} More formally, we try to find $v_k$, $k=2, \dots, J$, characterized by the following update, for  $k=2, \dots, J $:

$$\left\{\begin{array}{lll}
	 \pi_1^1=0,&\displaystyle \pi_k^1 =\pi_k+v_k, & \text{with probability } 1-\pi_1\\
	 \pi_1^2=1,&\displaystyle \pi_k^2 =\pi_k-\frac{1-\pi_1}{\pi_1}v_k, & \text{with probability } \pi_1 
\end{array}
\right.$$
where $\pi_k^1$ are the updated inclusion probabilities if the decision is to omit the current unit $k$, in this case $k = 1$, same for $\pi_k^2$ if we select the current unit $k$. The balancing equations need to be satisfied, we obtain the following equality:
$$
\sum_{k=1}^J \frac{\xb_k}{\pi_k}\pi_k^1 =\sum_{k=1}^J \frac{\xb_k}{\pi_k}\pi_k^2  =\sum_{k=1}^J\xb_k,
$$
that implies
$$
\sum_{k=1}^J \frac{\xb_k}{\pi_k}\pi_k^1 = 0+\sum_{k=2}^J \frac{\xb_k}{\pi_k}(\pi_k+v_k)
=\frac{\xb_1}{\pi_1}+\sum_{k=2}^J \frac{\xb_k}{\pi_k}\left(\pi_k-\frac{1-\pi_1}{\pi_1}v_k\right)
=\sum_{k=1}^J \frac{\xb_k}{\pi_k}\pi_k,
$$
and
$$
\sum_{k=2}^J \frac{\xb_k}{\pi_k}v_k
=\frac{\xb_1}{\pi_1}-\frac{1-\pi_1}{\pi_1}\sum_{k=1}^J \frac{\xb_k}{\pi_k}v_k
= \xb_1,
$$
and thus
$$
\sum_{k=2}^J \frac{\xb_k}{\pi_k}v_k= \xb_1.
$$
To ensure that updated inclusion probabilities remain between 0 and 1, we must also have
$$ 0 \leq \pi_k^1, \pi_k^2 \leq 1, \text{ for }k = 2,\dots,J, $$
which induces the following constraints:
\begin{equation}\label{eq:const}
	\max\left\{ -\pi_k, (\pi_k-1)\frac{\pi_1}{1-\pi_1}  \right\}\le v_k \le \min\left(  1-\pi_k,\pi_k\frac{\pi_1}{1-\pi_1} \right),k=2,\dots, J.
\end{equation}

In order to find $v_k$, we propose to solve the following program:
\begin{equation}\label{eq:linpro}\left\{ \begin{array}{ll}
	\displaystyle \text{maximize} & \displaystyle \sum_{k = 2}^J v_k c_k \\
	\text{subject to} & \displaystyle \sum_{k = 2}^J\frac{\xb_k}{\pi_k}v_k = \xb_1,\\
	& v_k \ge -\min\left\{ \pi_k, (1-\pi_k)\frac{\pi_1}{1-\pi_1}  \right\}, \\
	& v_k \le \min\left(  1-\pi_k,\pi_k\frac{\pi_1}{1-\pi_1} \right),
\end{array}\right.\end{equation}
where $c_k$ is a cost function that is supposed to penalized more the units that are far to current one. A naive function is $c_k = (J-k)$ which decreases as $k$ increases. If there is no solution the value of $J$ is incremented by one and we recompute until a solution is found. \textcolor{black}{The existence of the solution of the linear program \eqref{eq:linpro} is not guaranteed, mainly because if $J$ is not large enough, the balancing constraints and the positivity of the updated inclusion probabilities cannot be satisfied jointly. But as $J$ increases the feasible region expands.} The initial value for the variable $J$ must be larger than the dimension of $\xb_k$. If we let $\xb_k=\pi_k$, then the method only respects inclusion probabilities and we obtain fixed sample size.

\begin{figure}[htb!]
	\centering
	\includegraphics{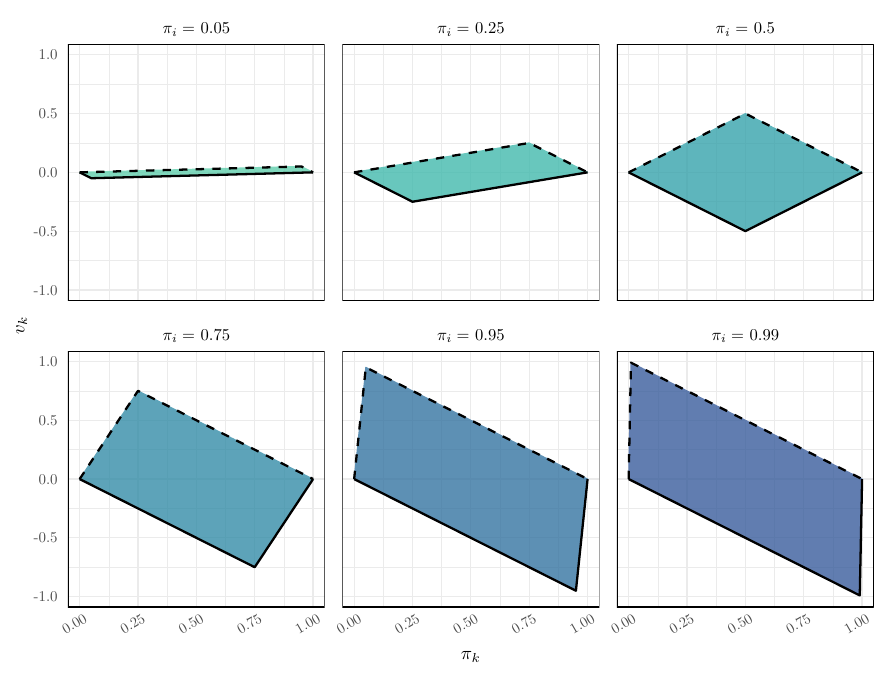}
	\caption{For six different inclusion probabilities $\pi_i$, we let vary $\pi_k$ from 0 to 1. On the $y$-axis, the bounds of $v_k$ in Equations \eqref{eq:const} are calculated. The coloured area represents the eligible value of $v_k$, the shaded line is the upper bound while the bottom line is the lower bound.}
	\label{fig:const}
\end{figure}

Suppose now that some steps have been proceed and that the current step is the $i$th one. Let denote $P_i = \{\pi_i,\dots,\pi_{i + M}\}$ the pool of available units at step $i$ where $M\in \N$. \textcolor{black}{Note that size of the pool $M$ is not directly related to $J$. The pool $P_i$ is actually all the available units for which we have waited to make a decision in the previous $i-1$ steps.} The algorithm will choose a unit in $P_i$ and make a decision on a certain amount of units $J$, with $J \leq M$. The constraint on the $v_k$ depends on the $\pi_i$, i.e., the inclusion probability of the current unit $i$ might have a serious impact on the potential value that could take $v_k$. Figure \ref{fig:const} shows possible $v_k$ against $\pi_k$, for different values of $\pi_i$. The value $\pi_k$ varies from 0 to 1 and the lower and upper bounds of the value $v_k$ are plotted. It shows clearly that when $\pi_i$ is very small, it ends with a tight interval for $v_k$. On the other hand, as the inclusion probabilities $\pi_i$ increase, we obtain a larger range for $v_k$. If inclusion probability of the current unit $i$ is very small, we might have a sharp increase of the value $J$. This comes from the fact that the balancing equations are not symmetric for unequal probability sampling. If a balanced sample is selected with unequal inclusion probabilities, the complementary sample is not a balanced sample.

To optimize the problem of sharp increase of the value $J$, we propose to always select as considered unit the one that has the largest inclusion probability in the pool $P_i$ of available units. By taking the one that has the largest inclusion probability and reordering with respect to the latter, we optimize the size of $J$.

If spatial coordinates are available, the proposed method could be used in an interesting way to select a well-spread sample. Let $P_{(i)} = \{\pi_{(i)}, \dots, \pi_{(i+M)}\}$, such that $\pi_{(i+j)}$ is the $j$th closest unit to $\pi_{(i)}$. Pool is so reordered with respect to the distance to the unit that has maximum inclusion probability. Doing that will then modify primarily inclusion probabilities of units closer to the current one. If a window has a sum of inclusion probabilities close to 1 and if the decision of the $i$th unit is to select it (i.e. inclusion probability is transformed to 1), it will ensure that nearest neighbours of the $i$th unit will not be selected. This leads to a selection process that spread very well the sample in the considered space.

If population units come from a stream, it is possible that some units have not yet appeared and that their spatial coordinates are in the neighbourhood of the $i$th unit. This side effect cannot be controlled, because, if we do not control which unit appears in the stream, it is not possible to know whether a future unit might appear within the considered window.


The method is comparable to the doubly balanced cube method \citep{gra:til:13}. Depending on the auxiliary variables, it might possible that the method does not end up directly with a sample. Some units might remain with inclusion probabilities not equal to an integer. In this case, we propose to land the method by launching the same process proposed in the landing phase of the doubly balanced method or cube method, either by suppression of variables or by linear programming. See Algorithm~\ref{algo:1} to have all details of the implementation.

\begin{algorithm}[htb!]
	\caption{Algorithm of sequential balanced sampling}\label{algo:1}
	Let $\pib$ be the inclusion probability vector, $\zb_k\in \R^q$ the spatial coordinates and $\xb_k\in \mathbb{R}^p$ the auxiliary variables of the $k$th unit. For $i = 1,2,\dots$, suppose a pool of units $P_i = \{\pi_i,\dots, \pi_{i + M}\}$.
	\begin{enumerate}
		
		\item Find $\pi_{(i)} \in P_i$ the maximum inclusion probability value in the pool. Define $P_{(i)} = \{\pi_{(i)}, \dots, \pi_{(i + M)} \}$ the pool of units reordered with respect to the distance of $\pi_{(i)}$ calculated using the spatial coordinates $\zb_k$, i.e. $\pi_{(i+j)}$ is the $j$th closest unit to the unit $\pi_{(i)}$.
		\item Find $J$ and $v_k,~ k = i+1,\dots, J$ such that the following linear program has a solution:
		$$\left\{ \begin{array}{ll}
			\displaystyle \text{maximize} & \displaystyle \sum_{k = i+1}^J v_k(J-(k-i)) \\
			\text{subject to} & \displaystyle \sum_{k = i+1 }^J\frac{\xb_{(k)}}{\pi_{(k)}}v_k = \xb_{(i)}\\
			& v_k \ge -\min\left\{ \pi_{(k)}, (1-\pi_{(k)})\frac{\pi_{(i)} }{1-\pi_{(i)}}  \right\}, \\
			& v_k \le \min\left(  1-\pi_{(k)},\pi_{(k)}\frac{\pi_{(i)}}{1-\pi_{(i)}} \right)
		\end{array}\right.$$
		At each step $J$ increases and units are reordered with respect to the distance of the current unit.
		\item Inclusion probabilities are modified on $P_{(i)}$. For $k = i+1,\dots, J$, let $q = \pi_{(i)}$,
		$$\left\{\begin{array}{llll}
			\pi_{(i)} = 0 &\text{and} & \pi_{(k)} = \pi_{(k)} + v_k & \text{with probability } 1 - q, \\
			\pi_{(i)} = 1 &\text{and} &\displaystyle \pi_{(k)} = \pi_{(k)} - \frac{1-\pi_{(i)}}{\pi_{(i)}} v_k & \text{with probability } q.
		\end{array}\right. $$
		The decision is then taken for unit $\pi_{(i)}$.
		\item The current unit $\pi_{(i)}$ and units that have inclusion probabilities transformed to an integer are removed from the pool and new units are included.
		\item Repeat steps 1-4 until it is no longer possible to find a solution to the linear program or if all inclusion probabilities are modified to an integer value.
		\item In general, it remains inclusion probabilities not equal to 0 or 1. A landing phase is launched by using either the doubly balanced sampling algorithm or landing phase of the cube method if there is no spatial coordinate.
	\end{enumerate}
\end{algorithm}

{\color{black}
\section{Variance Estimation}\label{sec:var}
Variance estimation of the Horvitz-Thompson estimator \eqref{eq:HT} requires the second-order inclusion probabilities. In general these probabilities are not known, therefore different variance estimators have been proposed. For maximum entropy design such as conditional Poisson sampling, an appropriate estimator is the H\'ajek-Ros\'en estimator
\begin{equation}\label{eq:varhaj}
  \widehat{\text{var}}_{HAJ}(\widehat{Y}) = \frac{n}{n-1} \sum_{k\in S}(1-\pi_k)\left\{ \frac{y_k}{\pi_k} - \frac{\sum_{\ell\in S} y_\ell (1 - \pi_\ell )/\pi_\ell}{\sum_{\ell\in S} (1-\pi_\ell)}  \right\}^2.
\end{equation}

 \citet{dev:til:05} show that the variance for balanced sampling methods can be computed as conditional variance with respect to balancing constraints. In particular they propose a general formula for variance approximation,

\begin{equation*}
  \text{var}_{app}(\widehat{Y}) = \sum_{k\in U}\sum_{\ell \in U} \frac{y_k}{\pi_k}\frac{y_\ell}{\pi_\ell} \Delta_{k\ell},
\end{equation*}
where 
$$ \Delta_{k\ell} = \left\{ \begin{array}{lll}
  \displaystyle b_k - b_k\ab_k^\top\left(  \sum_{i\in U} b_i\ab_i\ab_i^\top\right)^{-1} \ab_k b_k &~~ k = \ell \\
  \displaystyle -b_k\ab_k \left(  \sum_{i\in U} b_i\ab_i\ab_i^\top \right)^{-1} \ab_\ell b_\ell &~~ k \neq \ell.
\end{array}\right. $$

Different values for the parameter $b_k$ can be chosen. To obtain exact variance of the simple random sampling without replacement of fixed sample size, we must have $b_k = \pi_k(1 - \pi_k)\frac{N}{N-p}$. This approximated variance, calculated on the population using the same formula on the random sample $S$, gives an estimator of the variance:
\begin{equation}\label{eq:varest}
  \widehat{\text{var}}_{BAL}(\widehat{Y}) = \sum_{k\in S}\sum_{\ell \in S} \frac{y_k}{\pi_k}\frac{y_\ell}{\pi_\ell} \widehat{\Delta}_{k\ell},
\end{equation}
where 
$$ \widehat{\Delta}_{k\ell} = \left\{ \begin{array}{lll}
  \displaystyle c_k - c_k\ab_k^\top\left(  \sum_{i\in S} c_i\ab_i\ab_i^\top\right)^{-1} \ab_k c_k &~~ k = \ell \\
  \displaystyle - c_k\ab_k \left(  \sum_{i\in S} c_i\ab_i\ab_i^\top \right)^{-1} \ab_\ell c_\ell &~~ k \neq \ell
\end{array}\right. $$
and $c_k = (1-\pi_k)\frac{n}{n-p}$. Equation \eqref{eq:varest} can be rewritten by using residuals of linear model,
$$ e_k = y_k - \xb_k^\top \widehat{\betag} $$
where 
$$\widehat{\betag} = \left\{ \sum_{\ell \in S} (1-\pi_\ell) \ab_\ell \ab_\ell^\top \right\}^{-1}\sum_{\ell \in S} (1-\pi_\ell) \ab_\ell \frac{y_\ell}{\pi_\ell},$$
giving another expression for the Equation \eqref{eq:varest},
$$   \widehat{\text{var}}_{BAL}(\widehat{Y}) =  \frac{n}{n-p}\sum_{k\in S} (1-\pi_k) \left(\frac{e_k}{\pi_k}\right)^2.$$
More details about variance estimation under balanced sampling designs can be found in \citet{dev:til:05}. Estimator \eqref{eq:varest} is not taking into account for the spatial structure of the population. \citet{gra:til:13} propose another estimator based on a combination of the variance estimator \eqref{eq:varest} and the purely spatial variance estimator proposed by \citet{Stev:Olse:vari:2003},

\begin{equation}\label{eq:vardbs}
  \widehat{\text{var}}_{DBS}(\widehat{Y}) = \frac{n}{n-p}\frac{p+1}{p} \sum_{k \in S} (1-\pi_k)\left( \frac{e_k}{\pi_k} - \bar{e}_k \right)^2,
\end{equation}
where 
$$\bar{e}_k = \frac{\sum_{\ell \in G_k}(1-\pi_\ell) e_l/\pi_\ell }{\sum_{\ell\in G_k} (1-\pi_\ell)}$$
and $G_k$ is the set of the $p+1$ closest units of $k$ in the sample. 
\citet{gra:sch:14} developed also a generalized local mean variance estimator 
\begin{equation}\label{eq:varsb}
  \widehat{\text{var}}_{SB} (\widehat{Y}) = \frac{1}{2}\sum_{k\in S} \left( \frac{y_k}{\pi_k} - \frac{y_{\ell_k}}{\pi_{\ell_k}} \right)^2,
\end{equation}
where $\ell_k$ is the nearest neighbour to the unit $k$ in the random sample $S$. This expression of the estimator is a simple version when no equal distance exists. In its more general expression, the estimator can handle equal distance between units in the population and is showed appropriate for purely spatial sampling design such as local pivotal method \citep{gra:lun:sch:12}, proportional within distance \citep{ben:pier:17} and weakly associated vectors \citep{jaus:till:2020}. The following Section is devoted to a complete analysis of these methods and estimators.

}

\begin{figure}[ht!]
	\centering
	\input{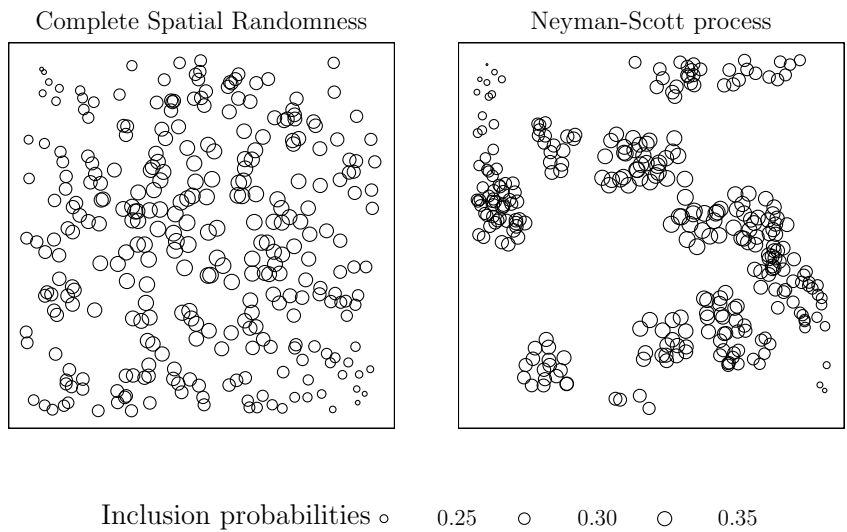}
	\caption{Simulated dataset used for the analysis of Section \ref{subsec:artificial}. The two datasets contain 300 units.}
	\label{fig:artificialExample2}
\end{figure}

{\color{black}

\section{Simulations}\label{sec:simu}

\subsection{Motivation on an artificial dataset}\label{subsec:artificial}

In this section, two artificial datasets are generated to analyze the performance of the proposed method. The two datasets are generated using functions of the package \texttt{spatstat} developed by \citet{spatstat}. The first dataset is generated from a complete spatial random process (CSR) using the function \texttt{rpoispp} while the second dataset is coming from a Neyman-Scott process (NS) using the function \texttt{rNeymanScott}. Figure \ref{fig:artificialExample2} shows the two datasets generated, the CSR process is completely random on the considered space while the NS is clustered. The population size $N$ of the two datasets is equal to 300.

Different methods are compared with the proposed method. Firstly, we compare with the doubly balanced sampling design proposed by \citet{gra:til:13}, as this method selects a well-spread balanced sample, the performance should be at least as good as this one. To see if we gain to incorporate the spatial structure of the population in the sampling method, we compare it with the cube method \citep{dev:til:04a,cha:til:06}. Next we compare with purely spatial sampling designs, namely, the local pivotal method \citep{gra:lun:sch:12}, the proportional within distance \citep{ben:pier:17}, the Halton iterative partitioning method \citep{rob:mcd:pri:brown:18} and the weakly associated vectors \citep{jaus:till:2020}. For each dataset two sets of inclusion probabilities are computed, one with equal inclusion probabilities and the other with unequal inclusion probabilities. Five auxiliary variables are generated using different probability distributions, a gaussian distribution $X_1 \sim \mathcal{N}(0,1)$, an exponential distribution $X_2 \sim \mathcal{E}(1)$, a gamma distribution $X_3\sim \Gamma(3,1)$, a beta distribution $X_4\sim \mathcal{B}(2,5)$ and an uniform distribution $X_5 \sim \mathcal{U}(0,3)$. In addition, a variable of interest $y$ is computed using the latter auxiliary variables,

\begin{equation}
    y = f(z_1,z_2) + X_1 + X_2 + X_3 + X_4 + X_5 + \mathcal{N}(0,1),
   \label{eq:y_artificial}
  \end{equation}
where 
$$ f(z_1,z_2) = 15 \text{exp}\left[ - \left\{ \left(z_1 - \frac{1}{2}\right)^2 - \frac{3}{5}\left(z_1 - \frac{1}{2}\right)\left(z_2 - \frac{1}{2}\right) - \left(z_2 - \frac{1}{2}\right)^2 \right\}  \right]. $$
The quantity $f(z_1,z_2)$ is a spatial autocorrelation computed from the density of a bivariate gaussian distribution. The variable $y$ is linear in the auxiliary variables. Figure \ref{fig:artificialExample2} shows the set of unequal inclusion probabilities proportional to the quantity $f(z_1,z_2)$. Table \ref{tab:dev_artificial} shows the relative deviation
\begin{equation}\label{eq:dev}
  RD^j = 100\times \frac{ \mid\widehat{t}_{X_i}(s) - t_{X_i}\mid}{t_{X_i}},
\end{equation}
where $t_{X_i}$ is the true total of the variable $X_i$, $i\in\{1,\dots,5\}$, $\widehat{t}_{X_i}(s) = \sum_{k\in s} X_{i,k}/\pi_k$ is the estimated totals of the auxiliary variables $\{X_1,\dots,X_5\}$ calculated on the sample $s$ which is a realization of the random sample $S$ specified by the sampling design, finally, $j$ is a notation to specify the sampling design used, for example $RD^{srs}$ for simple random sampling. The relative deviation shows how well the totals of the auxiliary variables are respected compared to the true totals. We observe that the three methods that select a balanced sample decrease drastically the relative deviation. All the methods respect the sample size as the first column is all equal to zero.

Let $Y$ denote the total of the variables $y$. For each sampling design, we compute the simulated variance based on $m$ simulations. Let denote this quantity by
\begin{equation}\label{eq:varsim}
  \text{var}_{sim}^j(\widehat{Y}) = \frac{1}{m}\sum_{s} \left\{\widehat{Y}_{HT}(s) - Y\right\}^2.
\end{equation}
To measure the accuracy of the method, we compute the ratio between the simulated variance of the sampling design and a reference simulated variance, namely, simple random sampling with fixed sample size for equal inclusion probabilities and conditional Poisson sampling for the set of unequal inclusion probabilities. We call this quantity the relative variance reduction denoted by
\begin{equation}\label{eq:varRV}
  RV^j = 100 \times \frac{\text{var}_{sim}^j(\widehat{Y})}{\text{var}_{sim}^{srs}(\widehat{Y})}.
\end{equation}
This quantity gives the percentage of variance reduction compared to the maximum entropy sampling design. In other words, it measures the accuracy of the sampling design. The more this value is small, the better we reduce the variance compared to the simple random sampling (respectively the conditional Poisson sampling). 

To see if the variance estimators discussed in Section \ref{sec:var} can retrieve the true variance, we compute the ratio between the variance estimator and the simulated variance. This quantity is named the relative efficiency of the variance estimator and is denoted by
\begin{equation}\label{eq:varRE}
  RE^j = 100 \times \frac{\widehat{\text{var}}^j(\widehat{Y})}{\text{var}_{sim}^j(\widehat{Y})}.
\end{equation}
Note that the variance estimator $\widehat{\text{var}}^j(\widehat{Y})$ depends on the sampling design $j$. For the proposed method and the doubly balanced sampling design, we use estimator \eqref{eq:vardbs} while for the local pivotal design, the proportional within distance, the Halton iterative partitioning and the weakly associated vectors methods, we use the variance estimator \eqref{eq:varsb}. Finally, for the simple random sampling design and conditional Poisson sampling design, we use the Hajek-Rosen estimator \eqref{eq:varhaj}.
From Table \ref{tab:tab_artificial} based on the results of the 10 000 simulations, we note that in terms of spread measures the proposed method is slightly better compared to the doubly balanced sampling design. Of course, a purely spatial sampling design gives a better spreading measure. But in terms of variance, the proposed method is the one that decreases the variance of the Horvitz-Thompson estimator the most compared to simple random sampling design or condition Poisson sampling. The variance estimators are showing acceptable performance except for unequal balanced sampling design. As the variable of interest is correlated to the balancing variables, the variance depends more on the rounding problem. The variance estimators discussed in Section \ref{sec:var} is known to fail to take into account the rounding problem. This effect is more discussed in \citet{leu:eus:jau:til:2021}. 
}

\begin{table}

\caption{\label{tab:tab_artificial}Results of 10 000 simulations on the variables of interest \eqref{eq:y_artificial}. The first column represent the relative variance reduction \eqref{eq:varRV}. The second columns contains relative variance estimator efficiency \eqref{eq:varRE}. The third and fourth columns correspond to the two spatial measures \eqref{eq:sb} and \eqref{eq:moran}.}
\centering
\resizebox{\linewidth}{!}{
\fontsize{11}{13}\selectfont
\begin{tabular}[t]{lrrrr}
\toprule
\multicolumn{1}{c}{ } & \multicolumn{1}{c}{Simulated Variances } & \multicolumn{1}{c}{Variance Estimators} & \multicolumn{2}{c}{Spread measures} \\
\cmidrule(l{3pt}r{3pt}){2-2} \cmidrule(l{3pt}r{3pt}){3-3} \cmidrule(l{3pt}r{3pt}){4-5}
  & $RV$ & $RE$ & $B$ & $I$\\
\midrule
\addlinespace[2em]
\multicolumn{5}{l}{\textbf{Neyman-Scott process}}\\
\addlinespace[1ex]
\multicolumn{5}{l}{Equal}\\
\hspace{1em}\hspace{1em}Proposed Method & 15.303 & 99.981 & 0.276 & -0.109\\
\hspace{1em}\hspace{1em}Doubly balanced & 20.427 & 74.581 & 0.316 & -0.070\\
\hspace{1em}\hspace{1em}Cube Method & 28.439 & 80.492 & 0.470 & -0.016\\
\hspace{1em}\hspace{1em}Local Pivotal & 75.684 & 112.498 & 0.145 & -0.328\\
\hspace{1em}\hspace{1em}Proportional within distance & 77.837 & 115.138 & 0.091 & -0.456\\
\hspace{1em}\hspace{1em}Halton Iterative Partitioning & 56.588 & 140.473 & 0.200 & -0.216\\
\hspace{1em}\hspace{1em}Wave & 77.830 & 109.194 & 0.145 & -0.524\\
\hspace{1em}\hspace{1em}Simple random sampling & 100.000 & 103.678 & 0.465 & -0.017\\
\addlinespace[1ex]
\multicolumn{5}{l}{Unequal}\\
\hspace{1em}\hspace{1em}Proposed Method & 6.976 & 2.680 & 0.277 & -0.115\\
\hspace{1em}\hspace{1em}Doubly Balanced & 8.567 & 2.188 & 0.317 & -0.070\\
\hspace{1em}\hspace{1em}Cube Method & 7.315 & 2.543 & 0.469 & -0.015\\
\hspace{1em}\hspace{1em}Local Pivotal & 98.123 & 110.137 & 0.149 & -0.323\\
\hspace{1em}\hspace{1em}Wave & 97.282 & 109.907 & 0.148 & -0.516\\
\hspace{1em}\hspace{1em}Conditional Poisson sampling & 100.000 & 99.177 & 0.473 & -0.015\\
\addlinespace[2em]
\multicolumn{5}{l}{\textbf{Complete Spatial Randomness}}\\
\addlinespace[1ex]
\multicolumn{5}{l}{Equal}\\
\hspace{1em}\hspace{1em}Proposed Method & 15.848 & 112.268 & 0.202 & -0.115\\
\hspace{1em}\hspace{1em}Doubly balanced & 22.457 & 77.951 & 0.237 & -0.068\\
\hspace{1em}\hspace{1em}Cube Method & 30.954 & 79.184 & 0.355 & -0.015\\
\hspace{1em}\hspace{1em}Local Pivotal & 83.738 & 108.654 & 0.109 & -0.312\\
\hspace{1em}\hspace{1em}Proportional within distance & 82.931 & 94.118 & 0.067 & -0.433\\
\hspace{1em}\hspace{1em}Halton Iterative Partitioning & 80.341 & 111.220 & 0.117 & -0.262\\
\hspace{1em}\hspace{1em}Wave & 81.762 & 112.643 & 0.099 & -0.487\\
\hspace{1em}\hspace{1em}Simple random sampling & 100.000 & 99.514 & 0.353 & -0.018\\
\addlinespace[1ex]
\multicolumn{5}{l}{Unequal}\\
\hspace{1em}\hspace{1em}Proposed Method & 8.696 & 1.682 & 0.202 & -0.117\\
\hspace{1em}\hspace{1em}Doubly Balanced & 9.126 & 1.599 & 0.240 & -0.064\\
\hspace{1em}\hspace{1em}Cube Method & 7.558 & 1.906 & 0.348 & -0.014\\
\hspace{1em}\hspace{1em}Local Pivotal & 92.683 & 107.609 & 0.110 & -0.309\\
\hspace{1em}\hspace{1em}Wave & 98.593 & 99.883 & 0.101 & -0.479\\
\hspace{1em}\hspace{1em}Conditional Poisson sampling & 100.000 & 97.904 & 0.349 & -0.014\\
\bottomrule
\end{tabular}}
\end{table}

\begin{table}

\caption{\label{tab:dev_artificial}Results of 10 000 simulations of the relative deviation \eqref{eq:dev} on the artificial dataset presented in Section \ref{subsec:artificial}.}
\centering
\resizebox{\linewidth}{!}{
\fontsize{11}{13}\selectfont
\begin{tabular}[t]{lrrrrr}
\toprule
  & $X_1$ & $X_2$ & $X_3$ & $X_4$ & $X_5$\\
\midrule
\addlinespace[2em]
\multicolumn{6}{l}{\textbf{Neyman-Scott process}}\\
\addlinespace[1ex]
\multicolumn{6}{l}{Equal}\\
\hspace{1em}\hspace{1em}Proposed Method & 72.004 & 4.689 & 1.984 & 2.005 & 2.176\\
\hspace{1em}\hspace{1em}Doubly balanced & 81.825 & 5.151 & 2.380 & 2.226 & 2.432\\
\hspace{1em}\hspace{1em}Cube Method & 75.079 & 4.383 & 2.103 & 2.050 & 2.318\\
\hspace{1em}\hspace{1em}Local Pivotal & 342.078 & 15.724 & 8.238 & 8.387 & 9.049\\
\hspace{1em}\hspace{1em}Proportional within distance & 364.628 & 14.762 & 8.306 & 7.886 & 9.003\\
\hspace{1em}\hspace{1em}Halton Iterative Partitioning & 306.922 & 14.245 & 10.358 & 7.924 & 7.620\\
\hspace{1em}\hspace{1em}Wave & 344.991 & 15.149 & 8.004 & 8.501 & 9.115\\
\hspace{1em}\hspace{1em}Simple random sampling & 335.823 & 15.640 & 8.086 & 8.117 & 9.015\\
\addlinespace[1ex]
\multicolumn{6}{l}{Unequal}\\
\hspace{1em}\hspace{1em}Proposed Method & 82.261 & 5.503 & 2.234 & 2.328 & 2.439\\
\hspace{1em}\hspace{1em}Doubly Balanced & 86.840 & 5.613 & 2.405 & 2.409 & 2.562\\
\hspace{1em}\hspace{1em}Cube Method & 75.682 & 4.807 & 2.131 & 2.101 & 2.376\\
\hspace{1em}\hspace{1em}Local Pivotal & 346.160 & 15.770 & 8.121 & 8.326 & 9.017\\
\hspace{1em}\hspace{1em}Wave & 348.579 & 15.658 & 8.396 & 8.331 & 9.003\\
\hspace{1em}\hspace{1em}Conditional Poisson sampling & 339.662 & 15.917 & 8.325 & 8.378 & 9.035\\
\addlinespace[2em]
\multicolumn{6}{l}{\textbf{Complete Spatial Randomness}}\\
\addlinespace[1ex]
\multicolumn{6}{l}{Equal}\\
\hspace{1em}\hspace{1em}Proposed Method & 70.456 & 3.457 & 1.902 & 2.019 & 2.426\\
\hspace{1em}\hspace{1em}Doubly balanced & 80.290 & 4.807 & 2.348 & 2.201 & 2.494\\
\hspace{1em}\hspace{1em}Cube Method & 75.396 & 4.468 & 2.090 & 2.032 & 2.301\\
\hspace{1em}\hspace{1em}Local Pivotal & 332.964 & 15.811 & 8.066 & 8.379 & 8.935\\
\hspace{1em}\hspace{1em}Proportional within distance & 325.536 & 15.990 & 7.955 & 7.997 & 9.319\\
\hspace{1em}\hspace{1em}Halton Iterative Partitioning & 306.923 & 11.961 & 8.593 & 7.502 & 9.828\\
\hspace{1em}\hspace{1em}Wave & 327.274 & 15.868 & 8.158 & 8.236 & 8.851\\
\hspace{1em}\hspace{1em}Simple random sampling & 337.471 & 15.548 & 8.035 & 8.270 & 8.964\\
\addlinespace[1ex]
\multicolumn{6}{l}{Unequal}\\
\hspace{1em}\hspace{1em}Proposed Method & 74.921 & 3.877 & 2.016 & 2.479 & 2.733\\
\hspace{1em}\hspace{1em}Doubly Balanced & 83.685 & 4.773 & 2.319 & 2.539 & 2.743\\
\hspace{1em}\hspace{1em}Cube Method & 75.100 & 4.367 & 2.083 & 2.197 & 2.434\\
\hspace{1em}\hspace{1em}Local Pivotal & 332.398 & 15.636 & 8.111 & 8.486 & 9.294\\
\hspace{1em}\hspace{1em}Wave & 330.964 & 15.606 & 7.995 & 8.477 & 9.123\\
\hspace{1em}\hspace{1em}Conditional Poisson sampling & 339.515 & 15.720 & 8.100 & 8.485 & 9.117\\
\bottomrule
\end{tabular}}
\end{table}

{\color{black}
\subsection{Real example on Amphibians dataset}\label{subsec:amphib}

In this section, the performance of the proposed method is analysed on a real dataset. The "Centre de coordination pour la protection des amphibiens et des reptiles de Suisse" makes available a dataset containing spatial coordinates and auxiliary variables on amphibians species. This dataset contains a list of 959 sites on which 19 species of amphibians are potentially observed. Figure \ref{fig:amphibian_data} shows two plots, the upper plot shows biogeographical regions of Switzerland while the lower plot shows the different sites. The size of the sites are displayed as well as the diversity score which is the count of the number of species observed on the site divided by the rarity score of the species. The rarity score is an ordinal variable that determines the scarcity level of the species \{1 = "endangered", 2 = "critically endangered", 3 = "threatened", 4 = "potentially threatened"\}.

Other auxiliary variables are available in the dataset. The ones used for our analysis are the altitude of the sites, the indicator variables of the biogeographical region and the area of the sites. Differents sampling designs are analysed, the first one uses equal inclusion probabilities while the unequal design uses inclusion probabilities proportional to the area variable. Table \ref{tab:dev_amphib} shows the relative deviation \eqref{eq:dev} on the auxiliary variables. Table \ref{tab:amphibian_tab} shows results of 10 000 simulations. As explained in Section \ref{subsec:artificial} performance of the proposed method is analysed using the ratio \eqref{eq:varRV} and \eqref{eq:varRE}. We note that the performance of the method is comparable to the doubly balanced sampling. The proposed method has good properties, it reduces the variance and shows good spread measures. We observe the same effect as the variance estimator fails to catch the rounding problem for unequal balanced sampling design.
}

\begin{knitrout}
\definecolor{shadecolor}{rgb}{0.973, 0.973, 0.973}\color{fgcolor}\begin{figure}
\includegraphics[width=\maxwidth]{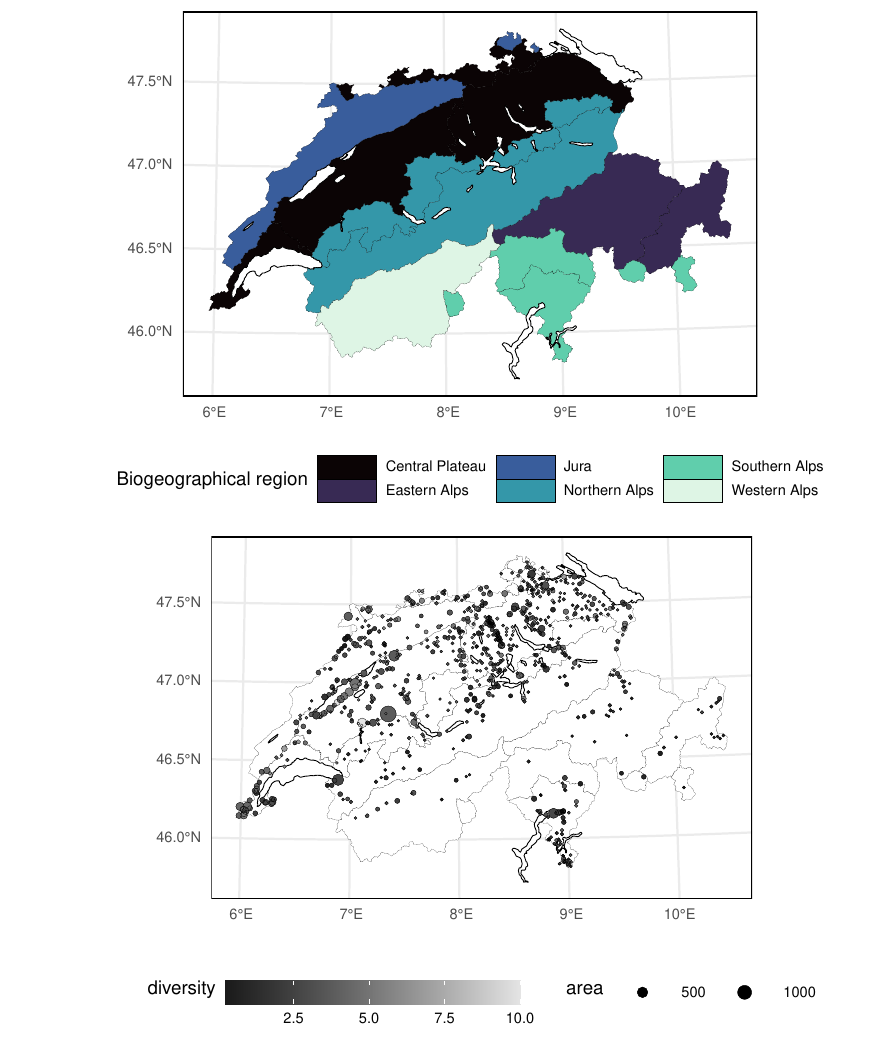} \caption[The upper plot display the biogeographical region of Switzerland while the lower plot gives the different sites of the amphibians dataset]{The upper plot display the biogeographical region of Switzerland while the lower plot gives the different sites of the amphibians dataset. The size of the point show the }\label{fig:amphibian_data}
\end{figure}

\end{knitrout}

\begin{table}

\caption{\label{tab:amphibian_tab}Results of 10 000 simulations on the diversity score of the amphibian dataset. The first column represent the relative variance reduction \eqref{eq:varRV}. The second columns contains relative variance estimator efficiency \eqref{eq:varRE}. The third and fourth columns correspond to the two spatial measures \eqref{eq:sb} and \eqref{eq:moran}.}
\centering
\resizebox{\linewidth}{!}{
\fontsize{11}{13}\selectfont
\begin{tabular}[t]{lrrrr}
\toprule
\multicolumn{1}{c}{ } & \multicolumn{1}{c}{Simulated Variance} & \multicolumn{1}{c}{Variance Estimator} & \multicolumn{2}{c}{Spread measures} \\
\cmidrule(l{3pt}r{3pt}){2-2} \cmidrule(l{3pt}r{3pt}){3-3} \cmidrule(l{3pt}r{3pt}){4-5}
  & $RV$ & $RE$ & $B$ & $I$\\
\midrule
\addlinespace[1ex]
\multicolumn{5}{l}{Equal}\\
\hspace{1em}Proposed Method & 71.376 & 106.592 & 0.280 & -0.160\\
\hspace{1em}Doubly Balanced & 69.231 & 110.391 & 0.196 & -0.243\\
\hspace{1em}Cube Method & 73.937 & 103.168 & 0.400 & -0.019\\
\hspace{1em}Local Pivotal & 70.240 & 119.056 & 0.132 & -0.387\\
\hspace{1em}Proportional within distance & 96.908 & 86.323 & 0.089 & -0.479\\
\hspace{1em}Halton Iterative Partitioning & 95.734 & 86.451 & 0.204 & -0.223\\
\hspace{1em}Simple random sampling & 100.000 & 95.826 & 0.435 & -0.005\\
\addlinespace[1ex]
\multicolumn{5}{l}{Unequal}\\
\hspace{1em}Proposed Method & 68.453 & 6.842 & 0.225 & -0.104\\
\hspace{1em}Doubly Balanced & 50.478 & 9.238 & 0.219 & -0.092\\
\hspace{1em}Cube Method & 38.209 & 12.221 & 0.397 & 0.014\\
\hspace{1em}Local Pivotal & 90.041 & 103.984 & 0.154 & -0.148\\
\hspace{1em}Conditional Poisson sampling & 100.000 & 98.740 & 0.407 & 0.016\\
\bottomrule
\end{tabular}}
\end{table}

\begin{table}

\caption{\label{tab:dev_amphib}Results of 10 000 simulations of the relative deviation \eqref{eq:dev} for the amphibians dataset presented in Section \ref{subsec:amphib}.}
\centering
\resizebox{\linewidth}{!}{
\fontsize{11}{13}\selectfont
\begin{tabular}[t]{lrrrrrrr}
\toprule
  & Area & Jura & Central Plateau & N Alps & W Alps & E Alps & S Alps\\
\midrule
\addlinespace[1ex]
\multicolumn{8}{l}{Equal}\\
\hspace{1em}Proposed Method & 0.765 & 4.068 & 0.352 & 3.208 & 33.531 & 11.482 & 5.085\\
\hspace{1em}Doubly Balanced & 0.795 & 4.057 & 0.352 & 3.219 & 33.883 & 11.379 & 5.149\\
\hspace{1em}Cube Method & 0.729 & 4.065 & 0.350 & 3.215 & 32.942 & 11.159 & 4.933\\
\hspace{1em}Local Pivotal & 2.832 & 10.403 & 2.731 & 8.255 & 42.388 & 17.321 & 5.691\\
\hspace{1em}Proportional within distance & 4.064 & 9.891 & 2.836 & 7.221 & 37.176 & 13.419 & 4.595\\
\hspace{1em}Halton Iterative Partitioning & 2.705 & 12.844 & 3.335 & 8.861 & 39.506 & 19.568 & 9.369\\
\hspace{1em}Simple random sampling & 4.510 & 20.969 & 6.096 & 17.747 & 66.740 & 41.436 & 24.637\\
\addlinespace[1ex]
\multicolumn{8}{l}{Unequal}\\
\hspace{1em}Proposed Method & 14.574 & 28.955 & 12.366 & 31.117 & 96.362 & 99.002 & 56.868\\
\hspace{1em}Doubly Balanced & 14.805 & 28.697 & 12.142 & 32.201 & 96.722 & 97.855 & 57.298\\
\hspace{1em}Cube Method & 14.423 & 27.578 & 10.995 & 29.972 & 98.300 & 98.165 & 56.205\\
\hspace{1em}Local Pivotal & 23.154 & 39.719 & 20.170 & 40.282 & 101.595 & 100.094 & 60.179\\
\hspace{1em}Conditional Poisson sampling & 24.605 & 44.360 & 22.206 & 46.113 & 110.398 & 111.400 & 68.156\\
\bottomrule
\end{tabular}}
\end{table}


	\section{Conclusion}\label{sec:conclu}
	
	The increasing amount of data in our century is going to make stream sampling very useful in few decades. In some experiments in environmental studies, it is even impossible to know the entire population and so sequential sampling are mandatory.
	In this manuscript, we proposed a completely new sequential method to select a balanced sample. This method respects equal and unequal inclusion probabilities. We showed through simulations that the proposed method is comparable with the doubly balanced sampling design in terms of variance of the Horvitz-Thompson estimator for a finite population.
	To conclude we can state that the method is an effective improvement for stream sampling and proposes a real enhancement in this field.

	
	\clearpage
	

\end{document}